\documentclass[doublecol]{epl2}

\usepackage{amsmath}
\usepackage{amssymb}
\usepackage{graphicx}
\usepackage{color}

\title{Kolmogorov analysis detecting radio and Fermi gamma-ray sources in cosmic microwave background maps}
\shorttitle{Kolmogorov analysis detecting radio and Fermi $\gamma$-ray sources in cosmic microwave background maps}

\author{V.G.Gurzadyan\inst{1} \and A.L.Kashin\inst{1} \and H.G.Khachatryan\inst{1} 
   \and A.A.Kocharyan\inst{1,2} 
   \and E.Poghosian\inst{1} 
   \and D.Vetrugno\inst{3}
   \and G.Yegorian\inst{1}
   }

\shortauthor{V.G.Gurzadyan \etal}

\institute{
\inst{1} Alikhanian National Laboratory and Yerevan State University, Yerevan, Armenia\\
\inst{2} School of Mathematical Sciences, Monash University, Clayton, Australia\\
\inst{3} Salento University and INFN, Sezione di Lecce, Lecce, Italy, EU
}


\abstract{The Kolmogorov stochasticity parameter is shown to act as a tool to detect point sources in the
cosmic microwave background (CMB) radiation temperature maps. Kolmogorov CMB map constructed for the WMAP's 7-year datasets reveals tiny structures which in part coincide with point radio and Fermi/LAT gamma-ray sources.
In the first application of this method, we identified several sources not present in the then available 0FGL Fermi catalog.  Subsequently they were confirmed in the more recent and more complete 1FGL catalog, thus strengthening the evidence for the power of this methodology.}

\begin{document}

\maketitle

%\label{firstpage}

%\begin{keywords}
%cosmic microwave background radiation.
%\end{keywords}

\section{Introduction}

A number of point sources are identified in the temperature maps of cosmic microwave background (CMB), including in the Wilkinson Microwave Anisotropy Probe's (WMAP) data \cite{Ja}; the point source catalog is available in http://lambda.gsfc.nasa.gov/product/map/. The efficiency of the search of point sources is particularly important for obtaining a pure cosmological signal of high precision at large multipoles, for the cross correlation studies with galaxy surveys, on baryonic oscillations, e.g. \cite{Saw1}\cite{Saw2}.    

We now involve for this aim a new descriptor, the Kolmogorov stochasticity parameter which describes the degree of randomness of a given sequence \cite{Kolm}\cite{Arnold}. It has been already been applied to the CMB temperature datasets \cite{KSP}. The resulting Kolmogorov CMB maps showed the potential of this approach in the separation of the Galactic disk from the cosmological signal, in the identification of anomalies such as the non-Guassian Cold Spot \cite{Cruz}\cite{anomaly4}. Certain spots and regions noticed in the K-map can be compared with the studies with other descriptors \cite{Ross}. 

Here we constructed and analyzed Kolmogorov's map for  WMAP's 7-year temperature data\footnote{This study initially was based on the WMAP's 5-year data and 7-year data were involved later upon their release: the results are coinciding absolutely.} and obtained a list of 12 regions of a degree scale with anomalously high value of Kolmogorov's function (see below). Among those regions, 6 coincide with sources given in the catalog of the WMAP's point sources, where they are identified as radio galaxies,  while all 12 regions coincide with gamma-ray sources discovered by Fermi satellite's Large Area Telescope (LAT) \cite{Fermi1}; 4 among these regions in both catalogs are overlapping mutually. 

In the first version of this study (arXiv:1002.2155v), when only 0FGL catalog of Fermi/LAT was available \cite{Fermi0}, there were 3 regions which have practically identical values both of the Kolmogorov's function and of the CMB temperature distribution but were absent in either of mentioned catalogs, i.e. of the radio and Fermi gamma sources. One might thus expect that they also are still unidentified sample of point sources. Indeed, the release of Fermi/LAT 1FGL catalog confirmed this prediction, i.e. all 12 CMB structures are identified in that catalog, thus showing the predictive power of the approach.

The detected sources include blazars/flat spectrum radio quasars (FSRQs) and active galaxy  \cite{Fermi1}, which are distinguished by their multi-wavelength emission, from radio to gamma rays, and due to their superluminous nature are efficient in probing the distant Universe. 

\section{The Kolmogorov analysis of the CMB map}

The Kolmogorov statistic and the stochasticity parameter \cite{Kolm}\cite{Arnold} is defined for a sequence $\{X_1,X_2,\dots,X_n\}$ represented in increasing order, so that the distribution function is $F(x) = P\{X\le x\}\ $. Together with the empirical distribution function 

\begin{equation}
F_n(x) = \left\{
\begin{array}{ll}
  0, & x<X_1; \\
k/n, & X_k\le x<X_{k+1}, k=1,2,\dots,n-1;\\
  1, & X_n\le x ,\\
\end{array}
\right.
\end{equation}

\noindent
the stochasticity parameter is defined as $\lambda_n=\sqrt{n} \sup_x|F_n(x)-F(x)|$.
For any continuous function $F$ $
\lim_{n\to\infty}P\{\lambda_n\le\lambda\}=\Phi(\lambda)\ ,$
where 
\begin{equation}
\Phi(\lambda)=\sum_{k=-\infty}^{+\infty}\ (-1)^k\ e^{-2k^2\lambda^2}\ ,\, \Phi(0)=0,\, \  \lambda>0\ ,\label{Phi}
\end{equation}
the convergence is uniform and $\Phi$ is independent on $F$ \cite{Kolm}.

\begin{figure*}[!htb]
\begin{center}
\includegraphics[scale=0.45]{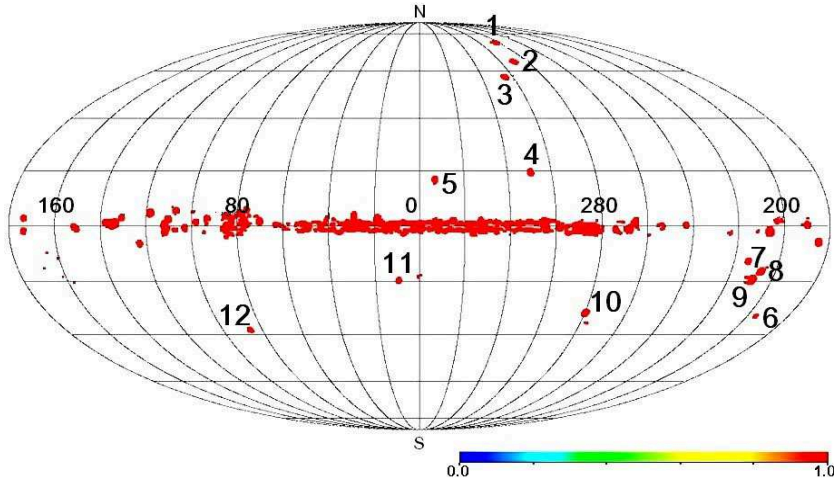}
\caption{\small{The location of the 12 high $\Phi$ regions in the sky, i.e. those outside 
the Galactic disk with $|b|>10^{\circ}$.}}
\label{f1.lbl}
\end{center}
\end{figure*}

The parameter $\lambda_n$ and the Kolmogorov's statistic define the degree of the randomness of the sequence, which can be tested via numerical experiments \cite{mod}. Namely, the stochasticity parameter gives a quantitative measure of the presence of correlations in a sequence of numbers independent of the underlying probability distribution within the interval of the probable values of $\lambda$ = 0.4-1.8.  This criterion was applied to sequences following from the number theory  \cite{Arnold}\cite{Arnold_ICTP}, and, as mentioned, to a physical problem, i.e. the CMB maps representing distribution of the temperature values over the set of pixels. The latter resulted in the distribution of Kolmogorov's function $\Phi$ over the sky which can reflect the properties of the large-scale matter inhomogeneities in the Universe i.e. on the distribution of the voids \cite{KSP}.  

The Kolmogorov's CMB maps were obtained for the Gaussian function for $F$ and for WMAP several w-band datasets in \cite{GK2009a} \cite{GK2009b}, where the procedure of averaging over the pixel sets in Healpix \cite{HP} representation is described.  {\it w}-band has the best resolution of all bands and is the least contaminated by the Galactic synchrotron radiation which is particularly important for us, as some of the studied regions are of not high Galactic altitude. In the distribution of $\Phi$, it appears that the regions with its values higher than 0.999 form a sample of 12 regions. Those 12 regions were found via two procedures: the $\Phi$ sky was scanned with a step $0.5^{\circ}$ with averaging of $\Phi$ within the regions of $1.5^{\circ}$ radii, and with a step $0.2^{\circ}$ and averaging within  $1.0^{\circ}$. The radius of the circles are determined by the available angular resolution of WMAP maps, i.e. at smaller regions the number of pixels becomes too small, while for larger regions the effect of the point source non-Gaussianity is smeared. The study of simulated maps in \cite{GK2009a}\cite{GK2009b} had revealed differences in the Kolmogorov's parameter in the WMAP's and generated pure Gaussian maps,
as the role of the WMAP's noise in the behavior and value of that parameter has been studied. The numerical simulations were performed also via generation of a given signal or several signals with the subsequent comparison of their $\Phi$-maps with the observed ones, in order to reveal the contribution of a given component. These studies had shown that Kolmogorov's parameter does reflect certain statistical properties of the CMB maps which enable the study of non-Gaussianities.  Direct quantitative comparison of the source detecting power of this approach and of other methods, however, is not much informative, since Kolmogorov's parameter is reflecting different type of correlations.    

The results of both scans revealed the same 12 regions at $|b|>10^{\circ}$ as shown in 
Fig. 1. Fig. 2 contains an example for the temperature and $\Phi$ behavior in one of those regions, No.04, of radius $5^{\circ}$.  In Fig. 3 we represent the CMB temperature behavior in the 12 regions denoted when radio (R), Fermi/LAT (G)  point source counterparts
are available. All $\Phi$ sources possess $\delta$-function type peak of temperature. Table 1 contains the coordinates of the revealed regions, the coordinates of the identified radio sources according to the WMAP's point source catalog, and Fermi/LAT gamma sources, along with the accuracy of identifications in arc minutes.  

\begin{figure*}
\begin{center}
\includegraphics[scale=0.3]{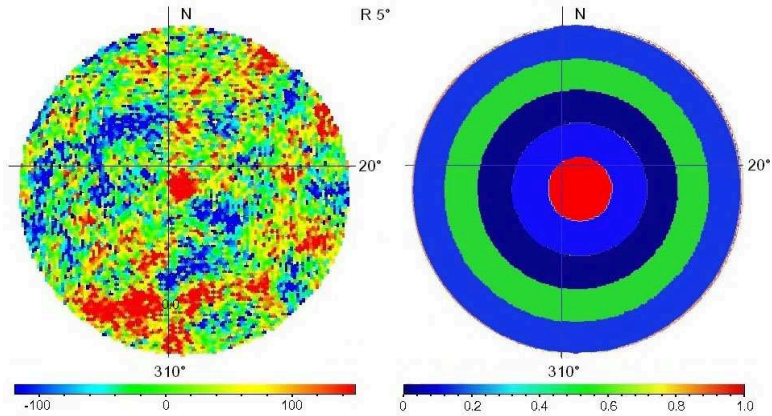}
\caption{\small{The CMB temperature map region of $5^{\circ}$ radius centered on No.04 (a), and its $\Phi$ distribution (b).}}
\label{f2.lbl}
\end{center}
\end{figure*}

\begin{figure*}
\begin{center}
\includegraphics[scale=0.36]{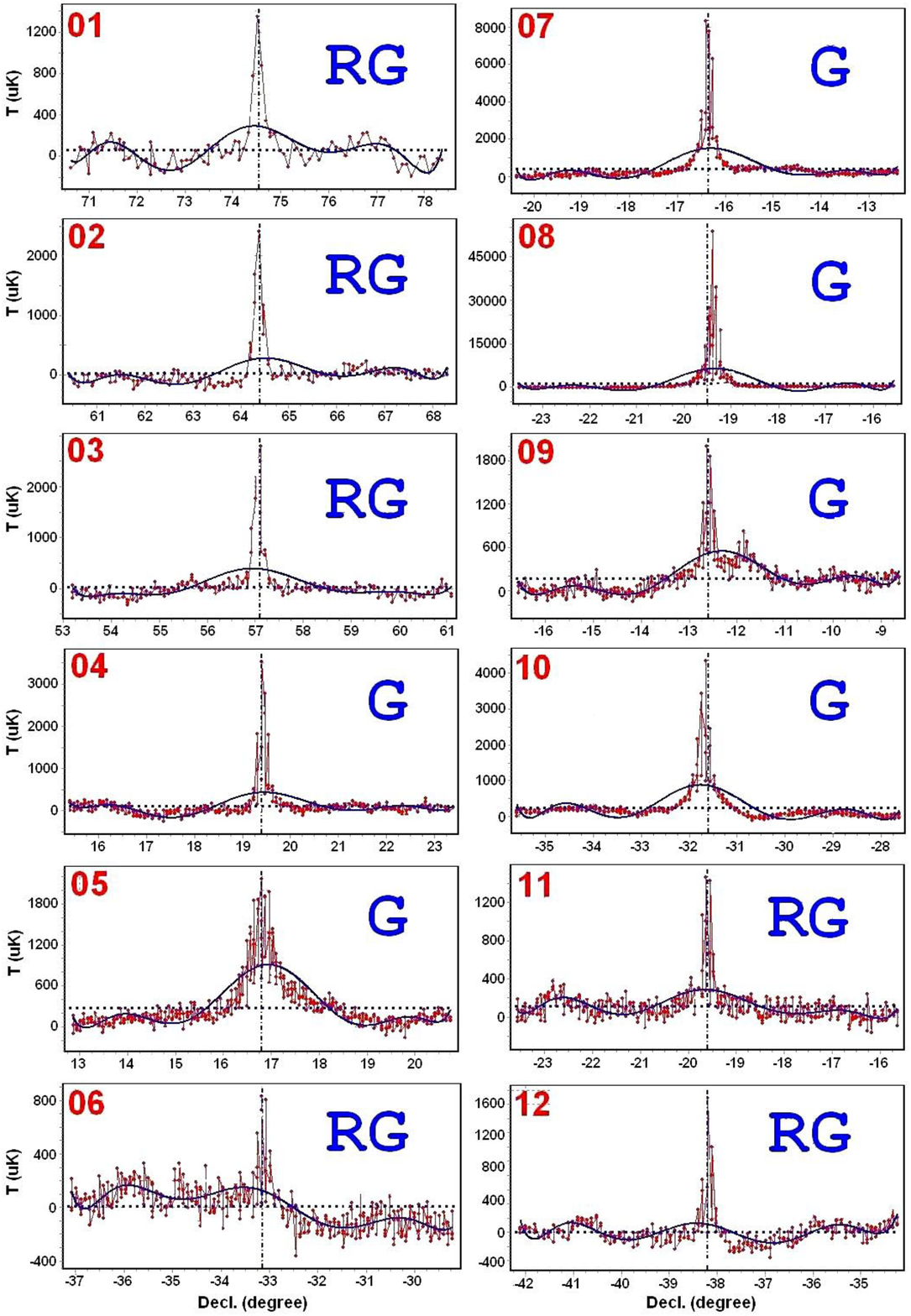}
\caption{\small{The distribution of the CMB temperature in the 12 regions (with smoothed, blue, line) ; (R) and (G) denote those with radio and Fermi/LAT gamma-source identifications, respectively.}}
\label{f3.lbl}
\end{center}
\end{figure*}

\section{Conclusion}

We used the Kolmogorov's parameter to extract point sources in the CMB maps. The idea is that the radiation of different origin can have different statistic and this can be revealed by the Kolmogorov's distribution. Numerical experiments for systems with built a priori  different statistic showed that K-parameter is indeed sensitive to the randomness properties of systems \cite{mod}.   Namely, the regions which might be of lower significance by conventional methods and hence are absent in WMAP's catalog, can be easily outlined by the present approach. On the other hand, the increase in the angular resolution of the CMB temperature maps will enable to apply this method for smaller scales, i.e. to obtain the averaged Kolmogorov function for large enough number of pixels, and hence to increase its source detecting power.

The analysis performed for WMAP's 7-year w-band maps aimed the extraction of around 1 degree scale regions (namely, when averaged within $1.0^{\circ}$ and $1.5^{\circ}$ circles at $0.2^{\circ}$ and $0.5^{\circ}$ step scans, respectively), which were outlined in the K-map due to their high value of the Kolmogorov's function $\Phi$. When compared with the catalog of the point sources of WMAP's corresponding map, 6 among those $\Phi$-regions were identified with radio galaxies, while all 12 regions were identified with Fermi/LAT gamma-ray 1FGL sources \cite{Fermi1}. The 3 detected anomal regions were absent in the previous 0FGL Fermi/LAT catalog \cite{Fermi0}.  However as concluded then, since the CMB regions have practically identical distributions of the function $\Phi$ as well as of the CMB temperature, the latter regions can also be point sources of close nature. Present study confirmed that prediction  and the interest for dedicated studies of CMB map's those regions in radio (see e.g. \cite{J}) and other bands. 

At the available angular resolution of WMAP the method enables to identify at high confidence level only a sample of 12 regions, which is certainly smaller than the number of sources identified by WMAP's mask. Again, the reason, as described above, is in the available resolution of the WMAP's maps. However, even at present resolution the method identifies sources not noticed by WMAP's procedure, which appear to be Fermi/LAT gamma sources \cite{Fermi1}. Most of those sources appear to be blazar/quasars and active galaxy, i.e. multi-wavelength and superluminous objects, and we showed another possibility of their identification based on totally different idea, the statistic of the signal.  The absence of some of the detected sources in the initial 0FGL catalog and their appearance in the next, 1FGL catalog, shows the predictive power of the approach. 

The Kolmogorov parameter thus enables to detect point sources in the CMB maps not noticed by other criteria.  The applicability of the method will increase at higher resolution maps expected from the PLANCK mission.

\begin{table*}
{
\renewcommand{\baselinestretch}{1.5}
\renewcommand{\tabcolsep}{1.3mm}
\small{
{\bf Table 1.} Coordinates of the 12 regions in WMAP's CMB map outlined by Kolmogorov's function $\Phi$, along with the coordinates of the identified radio sources, with the accuracy of identification in arcminutes dA and source designations, and the same for Fermi/LAT gamma-ray sources. \\[5pt]
}
\small{
\begin{center}
\begin{tabular}{crrrrcccrccl}
\hline 
\multicolumn{3}{c}{$\Phi$}& \multicolumn{4}{c}{Radio} &\multicolumn{4}{c}{Fermi/LAT}\\
No&\multicolumn{1}{c}{{\it l}}&\multicolumn{1}{c}{{\it b}} &\multicolumn{1}{c}{{\it l}}&\multicolumn{1}{c}{{\it b}}& D{\it A}& 
  &\multicolumn{1}{c}{{\it l}}&\multicolumn{1}{c}{{\it b}}& dA& 1FGL&nature \cite{Fermi1} \\[2pt]  
\hline
\hline
01&283.55& 74.55&283.81& 74.49&5.50&GB6 J1230+1223&283.83& 74.50& 5.36&J1230.8+1223 &active galaxy \\
02&289.90& 64.40&289.95& 64.36&2.73&GB6 J1229+0202&289.95& 64.36& 2.70&J1229.1+0203 &FSRQ \\
03&305.10& 57.10&305.11& 57.06&2.42&PMN J1256-0547&305.12& 57.06& 2.44&J1256.2-0547 &FSRQ \\
04&309.50& 19.40&      &      &    &              &309.55& 19.42& 3.34&J1325.6-4300 &active, radio galaxy \\
05&353.00& 16.80&      &      &    &              &353.47& 16.58&30.33&J1628.6-2419c&\\
06&195.35&-33.15&195.29&-33.14&3.07&PMN J0423-0120&195.25&-33.14& 5.30&J0423.2-0118 &FSRQ\\
07&206.50&-16.35&      &      &    &              &206.72&-16.38&12.82&J0541.9-0204c&\\
08&208.90&-19.50&      &      &    &              &209.07&-19.56&10.43&J0534.7-0531c&\\
09&213.70&-12.60&      &      &    &              &213.86&-12.58& 9.52&J0608.1-0630c&\\
10&279.60&-31.60&      &      &    &              &279.62&-31.63& 2.38&J0538.9-6914 &normal galaxy\\
11&  9.40&-19.60&  9.35&-19.61&2.89&PMN J1924-2914&  9.31&-19.72& 9.11&J1925.2-2919 &FSRQ\\
12& 86.00&-38.20& 86.12&-38.19&5.69&GB6 J2253+1608& 86.12&-38.19& 5.55&J2253.9+1608 &FSRQ\\
\hline
\end{tabular}
\end{center}
}}
\end{table*}

\end{document}